\documentclass[twocolumn]{aastex631}
\definecolor{mygreen}{RGB}{0, 185, 118}

\usepackage{amsmath, amssymb, amsfonts, latexsym}        % useful mathematical symbols

\begin{document}

\title{Observations of a New Form of Partial Filament Eruption}

\correspondingauthor{Abril Sahade}
\email{asahade@unc.edu.ar}

\author[0000-0001-5400-2800]{Abril Sahade}
\affiliation{Heliophysics Science Division, NASA Goddard Space Flight Center, Greenbelt, MD 20771, USA.}

\author[0000-0002-6975-5642]{Judith T. Karpen}
\affiliation{Heliophysics Science Division, NASA Goddard Space Flight Center, Greenbelt, MD 20771, USA.}

\author[0000-0003-0176-4312]{Spiro K. Antiochos}
\affiliation{CLaSP, University of Michigan, Ann Arbor, MI 48109, USA.}

\begin{abstract}
 Coronal mass ejections (CMEs) and coronal jets are two of the best-studied forms of solar eruptions, with the same underlying physics. Previous studies have presented partial eruptions producing coronal jets. We report, for the first time, a detailed analysis of three partial eruptions that segmented after the eruption began and produced CMEs. We use multiwavelength observations from the Solar Dynamics Observatory/Atmospheric Imaging Assembly, Solar TErrestrial RElations Observatory, and Solar Orbiter to reconstruct the three-dimensional evolution of the events. The magnetic field extrapolations indicated that the initial filaments were overlaid by pseudostreamer structures, and the splitting occurred after the interaction between the filament-supporting flux and external open field through their null points. The breakout mechanism seems to play a key role in both halting the system and splitting it. However, this initial evolution and consequent splitting into an erupting flux rope above the prominence segment that failed to erupt poses severe challenges to theories of solar eruptions.

\end{abstract}

\section{Introduction} 

Prominence/filament eruptions, flares, and coronal mass ejections (CMEs) 
are manifestations of the eruption of a stressed magnetic system \citep{Antiochos1998,Zhang2001,vanDriel2015,Green2018,Jiang2018,Yang2018,Filippov2019}. Filament channels, highly stressed magnetic structures that form
above polarity inversion lines (PILs) and store coronal free energy, are the
source regions of most eruptive events. Filament channels are force-balanced structures consisting of a highly sheared magnetic field (that is, nearly parallel to the PIL) restrained by an overlying unsheared field that links the opposite-polarity fluxes to either side of the PIL \citep{Martin1998}. While a magnetic flux rope (MFR) is largely accepted as the structure of the system once it erupts,  both sheared magnetic arcades (SMAs) and pre-existing MFRs are the most viable configurations for the pre-eruptive configuration of filament channels \citep[e.g.,][]{Patsourakos2020}. 

Knowing how much of a particular filament channel will erupt is, naturally, fundamental to predicting a CME event and the hazard it presents. Usually, we speak about full eruptions when most of the magnetic structure
escapes from the Sun, producing a CME, and about failed/confined eruptions when the eruptive process, including flares and filament activation, is halted in the low corona, with no magnetic structure escaping the Sun. However, there is a continuous transition between both cases, depending on the portion of the stressed magnetic system that is expelled.
Typically, these eruptions are denoted as partial filament eruptions and are the most frequently observed \citep{Gilbert2000,McCauley2015,Mateja2024}. 

Partial filament eruptions manifest as filament eruptions in which the initial filament splits into eruptive and non-eruptive segments. The initial filament can suffer a horizontal splitting, meaning that the segments belonged to different filament-channel sections along the PIL; or a vertical splitting, in which case both segments are located at different heights over the same section of the PIL. So far, horizontal splitting has been mainly interpreted as evidence for external reconnection of the filament-supporting stressed field and its surrounding field \citep[e.g.,][]{Zuccarello2009,Li2022,Kang2023}. On the other hand, vertical splitting has been attributed to the internal reconnection of a pre-eruptive MFR \citep[e.g.,][]{Liu2007,Tripathi2009,Kliem2014,Cheng2018}, or the existence of two magnetic systems in equilibrium in which only one destabilizes and erupts \citep[e.g.,][]{Liu2012,Cheng2014ApJ,Bi2015,Hou2023}.
Recent studies have shown that filaments can split by breakout reconnection in which part of the prominence material escapes through the reconnected overlying field lines, producing a jet, while the remaining part remains stable \citep{Zhang2022,Sun2023}.  In jet eruptions the filament channel reconnects with the ambient magnetic field \citep[e.g.][]{Wyper2017}, transferring helicity and plasma onto adjacent open field, so no MFR is expelled into the heliosphere. \citet{Mason2021} studied an event presenting also a partial eruption in a embedded dipole topology after the filament activation and resulting in a narrow CME-jet. The event represents a key link proving the connection on the magnetic nature of jets and CMEs, as end points of a broad spectrum of eruptive events \citep[see also,][]{Kumar2021,Wyper2024}.

In this work, we present three examples of partial filament eruptions in global magnetic topologies consistent with the breakout model \citep{Antiochos1999}, but that pose challenges for all theoretical models. The events were embedded in a pseudostreamer (PS) topology, which consists of a separatrix dome above a minority polarity region, and an outer spine emanating from a 3D null point on this dome that extends into the open heliosphere or connects to some distant closed-field region \citep[e.g.,][]{Wang2015,Raouafi2016,Mason2021,Wyper2021}. The events were simultaneously observed by the extreme ultraviolet (EUV) instruments onboard the \textit{Solar TErrestrial RElations Observatory Ahead} \citep[STEREO-A, STA hereafter,][]{STEREO_2008}, the \textit{Solar Dynamics Observatory} \citep[SDO,][]{SDO_2012SoPh}, and the \textit{Solar Orbiter} \citep[SolO,][]{SOLO_2020A&A}.  These cases differ from those analysed by \citet{Zhang2022} and \citet{Sun2023} in that the eruptive part became a CME instead of a jet, and differ from the presented by \citet{Mason2021} due to the filament partitioning during the eruption. Since the events analysed here were observed from multiple viewpoints, we are able to locate them through a three-dimensional (3D) reconstruction of the filament segments, and reconstruct their surrounding magnetic field.  We consider these examples representative of a larger category of partial eruptions in which the splitting occurs after the eruptive process started (filament activation), and one segment can be classified as a `successful' eruption (producing a CME) while the remaining segment becomes a `failed' eruption. 

The importance of these observations lies in the fact that they pose a severe challenge to all present models for solar eruptive events. To our knowledge, there are no numerical simulations reporting the type of partial eruptions presented below (in any of the different models available). From a theoretical point of view, the scenario in which an apparently coherent MFR forms and erupts upward but then splits into one MFR that continues to erupt as a CME flux rope and another one that stalls is difficult to explain. This issue will be discussed in more detail below. Section~\ref{sec:data} presents the instruments and methods used. Section~\ref{sec:analysis} presents the analysis and reconstruction of each event. Sections~\ref{sec:discussion} and \ref{sec:conclusions} present the discussion and summary of our analysis.

\section{Observations and methodology} \label{sec:data}

We study three well-observed events that allow us to analyze the filament splitting in different EUV filters and from different vantage points of view.
We use data provided by the \textit{Atmospheric Imaging Assembly} \citep[SDO/AIA,][]{AIA_2012SoPh} and wavelet-enhanced images \citep{Stenborg2008} from \textit{Extreme-Ultraviolet Imager }\citep[STA/EUVI,][]{SECCHI_2008}, and the \textit{Extreme Ultraviolet Imager} \citep[SolO/EUI,][]{EUI_2020A&A} to follow the prominence material and associated erupting plasma at higher temperatures. We also use white-light images from the \textit{Large Angle and Spectrometric Coronagraph Experiment} \citep[SOHO/LASCO,][]{LASCO_1995SoPh} and COR2 onboard STA to track the CME.

To understand the partitioning of the filament channels and their interactions with the magnetic environment, we tracked the real position of the encompassed prominence material.
We reconstruct the 3D evolution of the different segments of the filaments using the tie-pointing technique, which based on the epipolar constraint derives the position of a feature in 3D space when observed simultaneously in the fields of view (FOVs) of different spacecraft \citep[e.g.,][]{Inhester2006}.  We use here the \texttt{scc$\_$measure3} routine, adapted in \cite{scc_measure3} to include three viewpoints (available at \url{https://github.com/asahade/IDL}). Once the correspondence between a feature observed in image 1 and the projection of that feature observed in image 2 is identified, the 3D position of the feature is determined from the coordinates of the projected images. The position is then converted to the projection of the third image, which helps to validate or reject the selection \citep[see appendix of][]{Sahade2025}. The third image serves as a constraint, helping to eliminate ambiguity or ghost features (those unrelated to the material we are tracking). This technique is flexible enough to track plasma parcels (segments) of a filament without assuming a predetermined shape or model. 
To triangulate the position of different features, we look for images captured as close in time as possible. Since SolO has an elliptical orbit, its distance from the Sun varies considerably (from $\sim 0.3$ AU to more than $1$ AU); therefore, to pair the images, we use the header flag \textit{date\_ear}, which gives the light arrival time at 1 AU for a SolO observation. 

For each event with three viewpoints, we triangulated the position of different elements of the eruption to understand their evolution. 
We performed a careful reconstruction of different features with a cadence of ten minutes (SolO/EUI FSI304 cadence), and visually inspected the sequence of AIA 304 \AA\ images with their maximum cadence (36 seconds) to measure the displacements of the tracked features. 

Later in the evolution of the eruptions, we use the Graduated Cylindrical Shell (GCS) model to estimate the position of the ejected material in the coronagraph images. To reproduce the evolution of deflected CMEs from the low corona, we use a non-radial GCS model. In this way, the CME footpoint coordinates can be fixed according to the initial location of the eruptive segment, while the CME front can vary in latitude and longitude. We use the \textit{SolarSoft} routine \texttt{rtcloudwidget} with the following parameters tuned for the reconstruction: latitude, longitude, tilt angle, height, half angle, aspect ratio and non-radial tilt. The last allows us to adjust the angle subtended by the CME axis and the radial plane defined by the footpoints and the solar center, to better capture the CME front while the CME footpoints remain in the source region \citep[see, e.g.,][]{Sahade2023}.

For the coronal magnetic-field reconstruction we use the Potential Field Source Surface \citep[PFSS,][]{PFSS_2003SoPh} model. The PFSS model uses as input synoptic magnetograms updated every six hours with the central $120^\circ$ of data from the Helioseismic and Magnetic Imager \citep[SDO/HMI,][]{HMI2012}. Due to the location of the observed filaments (to the east limb of SDO), the source region magnetograms were old. However, the overlying topology remained stable throughout the Carrington rotation, so we decided to use the PFSS corresponding to the eruption date, except for the 2023 February 05 event. For the last one, we used the synoptic magnetogram updated few days later (on 2023-Feb-12) since the AIA observations indicated a closed loop system to the south better reproduced by the field lines corresponding to that date.

\section{Analysis and Results}\label{sec:analysis}
We present a detailed analysis of the observations in the different EUV filters from the three points of view. We tracked the 3D positions of the different segments of the partially erupted filaments to understand how the supporting field interacts with the surrounding magnetic field, to establish the moment of the splitting, and to follow their subsequent evolution.

\subsection{2022 October 28}
\begin{figure}    %%%%%%%%%%%%%%%%%% FIGURE 1 
   \centerline{\includegraphics[width=0.49\textwidth]{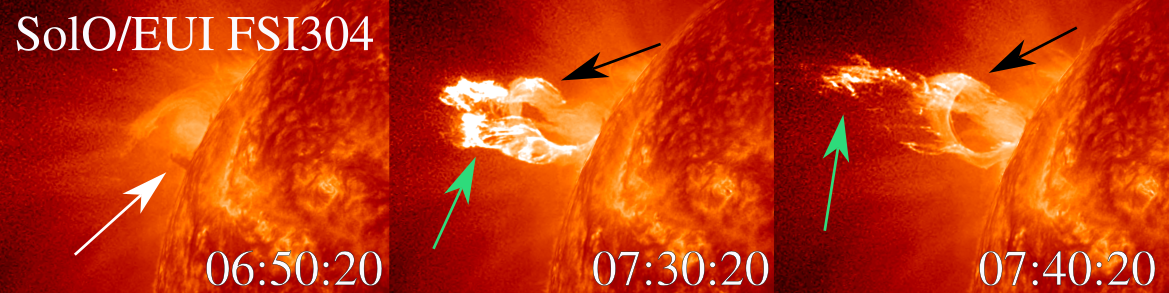}
             }
 \caption{Evolution of the splitting filament on 2022 October 28, seen by SolO/EUI FSI304. In the first panel, the white arrow indicates the rising filament. In the central and right panels, the green arrows indicate the eruptive segment, and the black arrows indicate the confined segment. The right panel shows the last stage of the separation, with brightenings also observed in hotter wavelengths. An animation showing the filament rising and splitting from 4:20 UT to 18:00 UT is available in the online version. The animation includes different views. The first, running from 0 to 8 seconds into the animation, is of the SolO/EUI FSI 304 {\AA} (left) and 174 {\AA} (right) evolution starting at 4:20 to 08:20. The second portion from 8 to 21 seconds is of the STA/EUVI 304 {\AA} (left) and base difference 195 {\AA} (right) evolution during the same period. Then, from 21 to 35 seconds, the SDO/AIA  94 {\AA} (top left), 131 {\AA} (top middle), 171 {\AA} (top right), base difference 193 {\AA} (bottom left), 211 {\AA} (bottom middle), and 304 {\AA} (bottom right) animation is presented. Last, from 35 to 39 seconds, the evolution from $\sim$4:20 to $\sim$17:50 in the FOVs of STA/COR2 and LASCO/C2 is displayed.}  
\label{fig:2022-10-28}
\end{figure}

The source of this partial eruption was a filament channel located on the northeast limb in the SolO/EUI FOV. Around 4:30 UT the filament started to rise slowly as one coherent structure. At 7:15 UT the filament showed brightenings at the splitting location in all EUV filters of AIA. The division was completed after 7:40 UT. Figure~\ref{fig:2022-10-28} shows three images from SolO/EUI in the 304 \AA\/ filter showing different stages of the evolution. The left image, at 6:50 UT, shows the rising filament before its separation (white arrow); the central image, at 7:30 UT, shows the separation between the eruptive and confined segments (green and black arrows, respectively); and the right image, 7:40 UT, shows the eruptive (green arrow) and confined (black arrow) segments of the filament once the splitting is complete. Note that both the eruptive and confined segments appear to share the same legs connecting to the chromosphere. This is clearly not a case of splitting along different sections of a PIL, as confirmed by the tracked position of the different features in time. Furthermore, there is no evidence of two disconnected filaments lying one above the other, as has been previously discussed in the literature \citep[e.g.,][]{Chen2021ApJ}.

The eruptive segment appeared in the LASCO C2 FOV at 8:12 UT as a CME. The average speed of the CME is about $900\,$ km/s, according to the GCS reconstruction, so it was a fairly fast CME. We have shown previously how a confined eruption can accompany a strong coronal jet, but in this case the eruption produced a clear CME. Somehow, the erupting segment was able to maintain its integrity and was not destroyed by interchange reconnection  within the FOV of the observations. The upper part of the confined segment showed a sideways motion until 8:05 UT, when it was no longer visible in the EUV filters.  Figure~\ref{fig:2022-10-28} is accompanied by an animation with the full evolution of the eruption within the different instruments. 

\begin{figure}    %%%%%%%%%%%%%%%%%% FIGURE 1 
   
   \centerline{\includegraphics[width=0.45\textwidth]{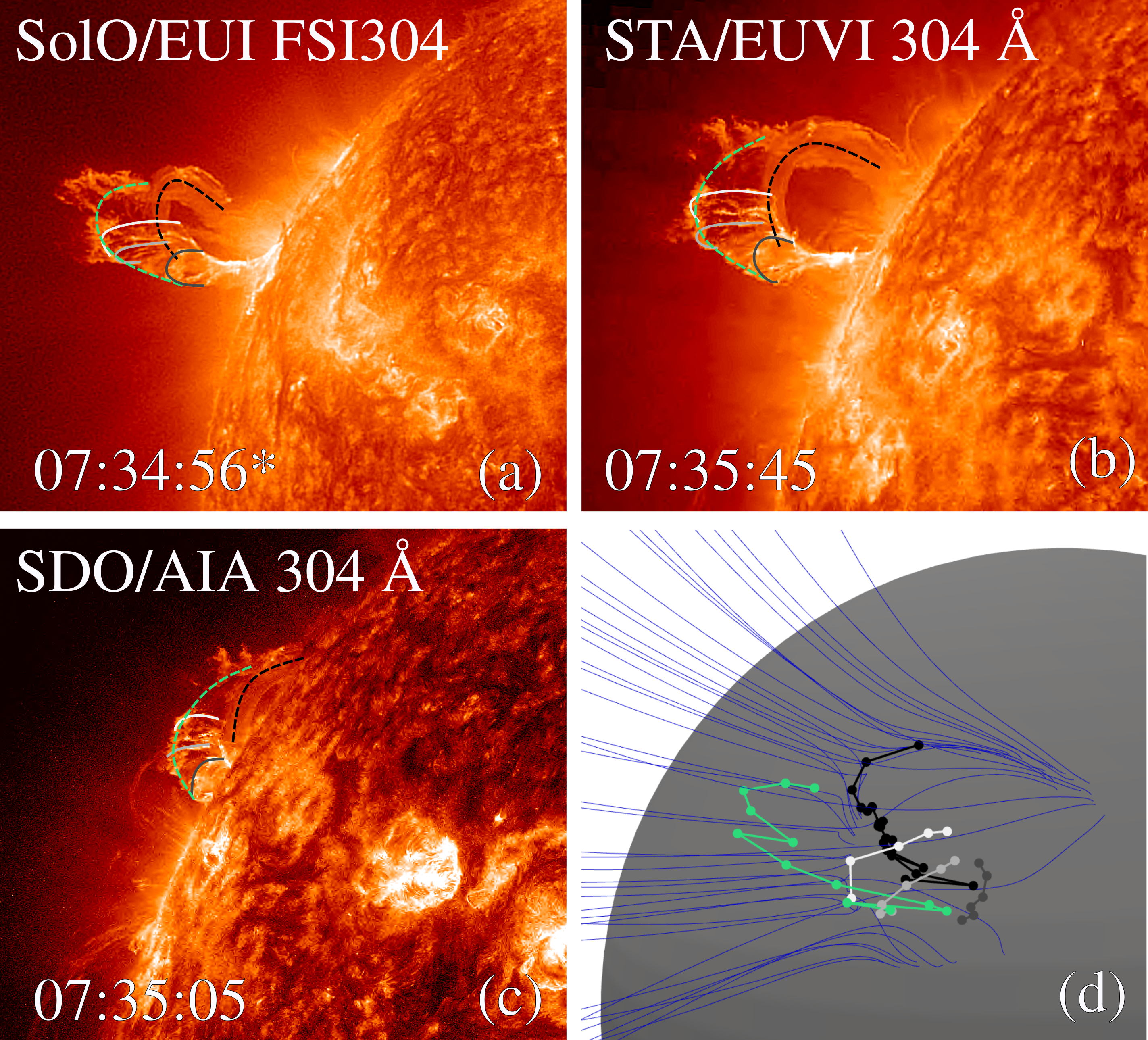} }
    
 \caption{Splitting filament on 2022 October 28 seen by SolO/EUI FSI304 (a), STA/EUVI 304 {\AA} (b), and SDO/AIA 304 {\AA} at 7:35 UT. SolO time corresponds to the time at Earth (\textit{date\_ear}) instead of the observation time, since the spacecraft received the image ahead at $d_\sun=0.44\,$AU. Schematically the reconstructed areas are indicated by lines projected in the different images. The green and black dashed lines indicate the eruptive and confined segments, respectively. The grey lines show the arcs connecting both segments. In panel (d) the 3D positions of the triangulated pixels are shown with the open field lines (blue) in the area of the eruption, and the colors of the segments as indicated above. The grey sphere represents the solar surface at $1\,R_\sun$, and to provide orientation the equator (teal), SolO meridian (yellow), STA meridian (pink), and SDO meridian (turquoise) are displayed. This figure is available online as an interactive figure, allowing different rotations and projections of the triangulated dots.  } 
\label{fig:3d}
\end{figure}

Figure~\ref{fig:3d} shows an example of the reconstructed features at 7:35 UT, in which we identify a lower arc corresponding to the confined segment (black line of each panel), an upper arc corresponding to the eruptive segment (green lines), and some arcs that seem to connect the eruptive and confined segments (grey arcs). Figure~\ref{fig:3d}(d) shows the 3D positions of the triangulated pixels and the open field lines surrounding the area from the PFSS reconstruction. The grey arcs apparently connect the eruptive and confined segments, indicating that the eruptive segment belonged to an outer envelope of the expanded filament. The confined segment (black dots) remained under the PFSS open field lines (below the PS null region), while the eruptive segment escaped throughout the null along a trajectory aligned parallel to the enveloping open field lines of negative polarity. An interactive version of Figure~\ref{fig:3d}(d) is available online, in which different views of the 3D reconstruction can be displayed. The CME related to the eruptive segment also showed a great deflection, aligning with the open field line direction.

Days after the eruption, the source region contained two filaments lying under the same PS, with the negative footpoints in the north and the positive footpoints in the south.  

\subsection{2023 February 5}

\begin{figure}    %%%%%%%%%%%%%%%%%% FIGURE 5 
   \centerline{\includegraphics[width=0.49\textwidth]{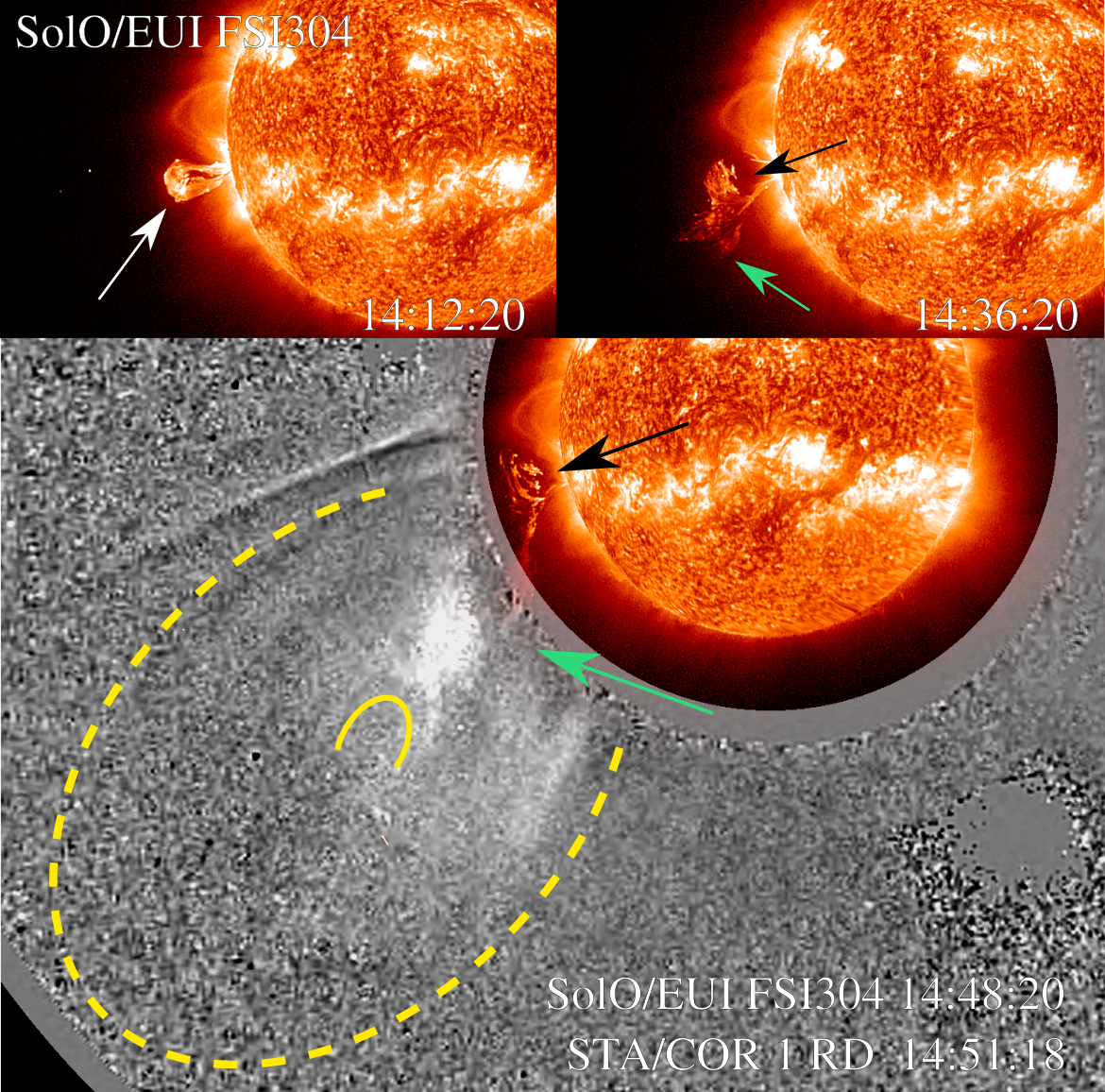}
             }
 \caption{Evolution of the splitting filament of 2023 February 5, as seen by SolO/EUI FSI304, and of the CME in a running-difference image from STA/COR 1. In the left upper panel the white arrow indicates the rising filament. In the right upper and lower panels the green arrows indicate the eruptive segment, and the black arrows indicate the confined segment. The yellow dashed and solid lines in the lower panel indicate the boundary of the CME and horn-like structure at the top part of the filament, respectively. An animation showing the filament rising and splitting from 11:00 UT to 18:00 UT is available in the online version. Details idem Fig.\ref{fig:2022-10-28}.
  An animation showing the filament rising and splitting from 11:00 UT to 18:00 UT is available in the online version. The animation includes different views. The first, running from 0 to 12 seconds into the animation, is of the SolO/EUI FSI 304 {\AA} (left) and 174 {\AA} (right) evolution. The second portion from 12 to 25 seconds is of the SDO/AIA  94 {\AA} (top left),131 {\AA} (top middle), 171 {\AA} (top right), base difference 193 {\AA} (bottom left), 211 {\AA} (bottom middle), and 304 {\AA} (bottom right) evolution. The third portion, from 25 to 35 seconds, is a high-cadence zoom-in evolution of SDO/AIA 171 {\AA} (top left), base difference 193 {\AA} (top right),211 {\AA} (bottom left), and 304 {\AA} (bottom right) during the splitting of the filament (14:00 to 15:00 UT). Last, from 35 to 44 seconds, the evolution from $\sim$11:00 to $\sim$18:00 in the FOV of STA/COR1 and LASCO/C2 is displayed.}
\label{fig:2023-02-05}
\end{figure}

\begin{figure}    %%%%%%%%%%%%%%%%%% FIGURE 6 
 
   \centerline{\includegraphics[width=0.45\textwidth]{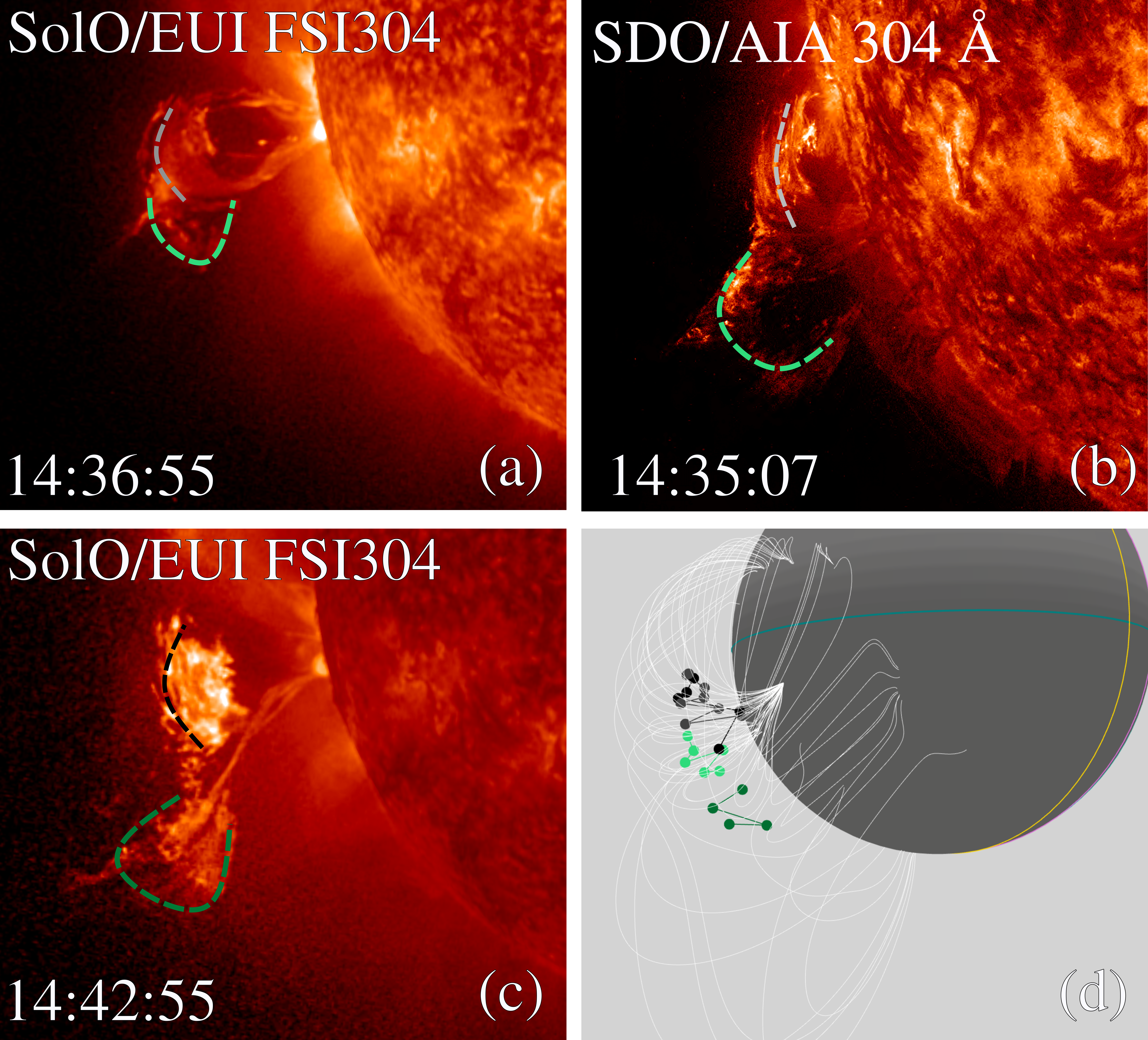} }
    
 \caption{Splitting filament of 2023 February 5 seen by SolO/EUI FSI304 (a), and SDO/AIA 304 {\AA} (b) at 14:35 UT. Panel (c) shows an image from SolO/EUI FSI304 at 14:42, when the splitting is advanced. Schematically the reconstructed areas are indicated by lines projected in the different images. In panels (a) and (b) the  lighter green and gray dashed lines indicate the eruptive and confined segments, respectively, corresponding to the same color dots of panel (d). The darker green and black dashed lines of panel (c) indicate the reconstructed area also displayed in panel (d) with dots of the same color. In panel (d) the 3D positions of the triangulated pixels at 14:35 and 14:42 UT are shown with the closed magnetic field lines (white) in the area of the eruption, with the segment colors indicated above. An interactive version of this figure is available online, which allows different rotations and projections of the triangulated dots. The interactive figure also includes negative-polarity open field lines surrounding the area, which are not displayed in the static figure to simplify the visualization. The grey sphere represents the solar surface at $1\,R_\sun$, and to provide orientation the equator (teal), SolO meridian (yellow), STA meridian (pink), and SDO meridian (turquoise) are displayed.  } 
\label{fig:3d_b}
\end{figure}

A filament located behind the southeast limb, as observed in the SolO/EUI FOV, partially erupted after 13:55 UT, when it was first observed by SolO/EUI FSI304 and FSI174. The filament ascended coherently, even forming a EUV front that is visible in FSI174 and the 94 {\AA} filter of SDO/AIA. The splitting process started at 14:24 UT, as shown in the high-cadence images from SDO/AIA. The confined segment disappeared at 15:01 UT, while the eruptive segment was first visible in LASCO C2 at 15:05 UT. The 304 {\AA} images show the west leg of the eruption, but the SolO/EUI FSI174 images reveal the nose of the eruptive segment and the preceding shock later observed by the coronagraph. Figure~\ref{fig:2023-02-05} shows in the top left panel the eruptive filament in the pre-splitting stage (white arrow) as observed by SolO/EUI FSI304; in the top right panel shows the erupting (green arrow) and confined (black arrow) segments, and also prominence material connecting both segments. In the bottom panel of Fig.~\ref{fig:2023-02-05} the eruptive segment (green arrow) is displayed nearly simultaneously in SolO/EUI FSI304 and STA/COR 1. The coronagraph image shows the CME shock and a horn structure ahead of the bright prominence material. Figure~\ref{fig:2023-02-05} includes a movie of the evolution from the different observatories and filters in the online version. The GCS reconstruction indicates a speed of $\sim 600\,$km/s.

\begin{figure}    %%%%%%%%%%%%%%%%%% FIGURE 1 
   \centerline{\includegraphics[width=0.49\textwidth]{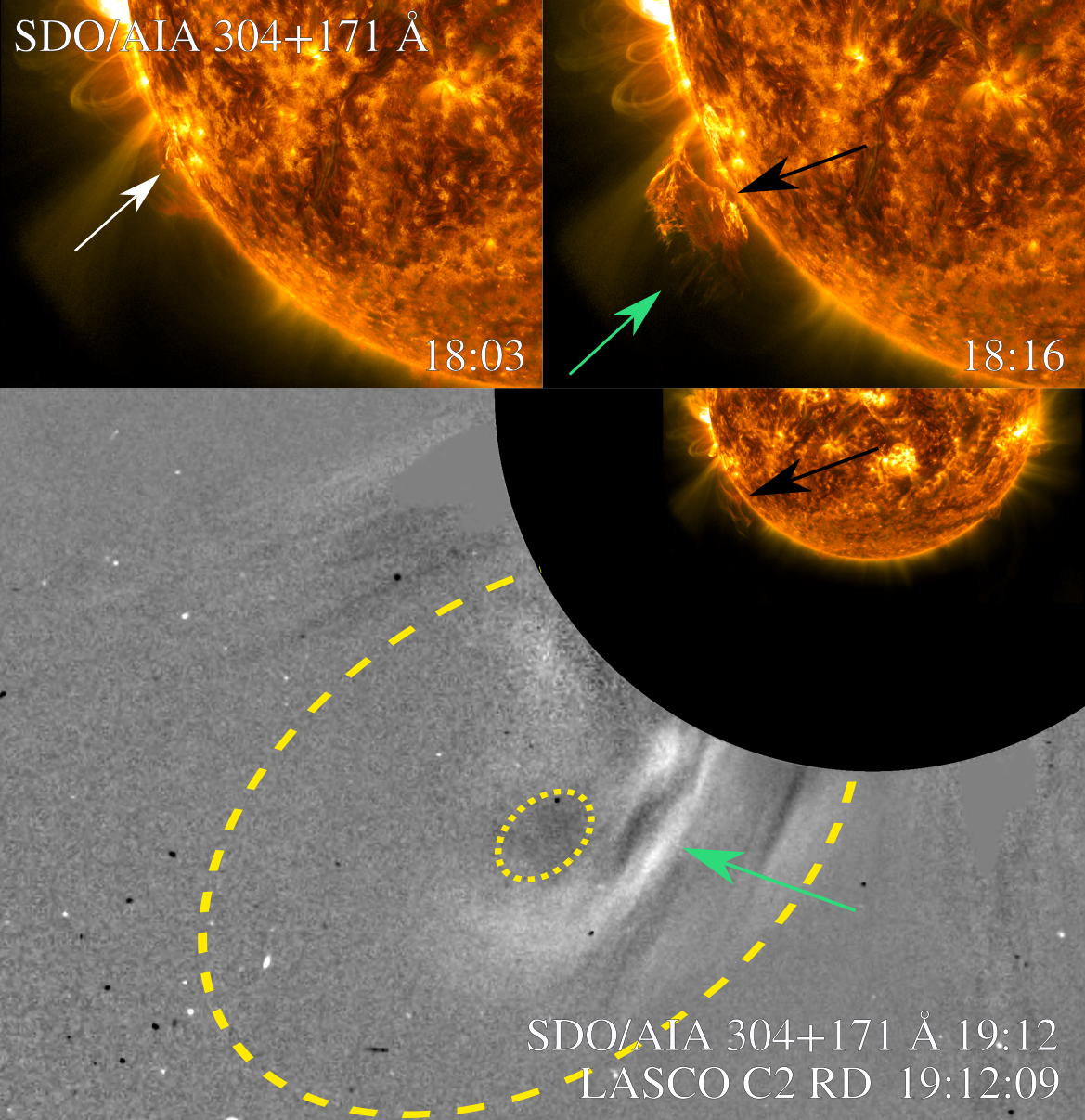}
             }
 \caption{Evolution of the event observed by SDO/AIA on 2023 October 15: composite images of the splitting filament in 304 and 171 {\AA} filters, and the CME running-difference image from LASCO C2. In the left upper panel the white arrow indicates the rising filament. In the right upper and lower panels the green arrows indicate the eruptive segment, and the black arrows indicate the confined segment. The yellow dashed and dotted lines in the lower panel indicate the front of the CME and the cavity inside. An animation showing the filament rising and splitting from 16:00 UT to 23:20 UT is available in the online version. The animation includes different views. The first, running from 0 to 8 seconds into the animation, is of the SolO/EUI FSI 304 {\AA} (left) and 174 {\AA} (right) evolution. The second portion from 8 to 26 seconds is of the STA/EUVI 304 {\AA} (left) and base difference 195 {\AA} (right) evolution during the same period. Then, from 27 to 44 seconds, the SDO/AIA  94 {\AA} (top left), 131 {\AA} (top middle), 171 {\AA} (top right), base difference 193 {\AA} (bottom left), 211 {\AA} (bottom middle), and 304 {\AA} (bottom right) animation is presented and a zoom-in high-cadence animation of the last four filters runs from 44 to 56 seconds of the video (from 17:50 to 18:35 UT). Lastly, from 56 to 62 seconds, the evolution from $\sim$16:00 to $\sim$23:10 in the FOV of STA/COR2 and LASCO/C2 is displayed.}
\label{fig:2023-10-15}
\end{figure}

\begin{figure}    %%%%%%%%%%%%%%%%%% FIGURE 1 
   
   \centerline{\includegraphics[width=0.45\textwidth]{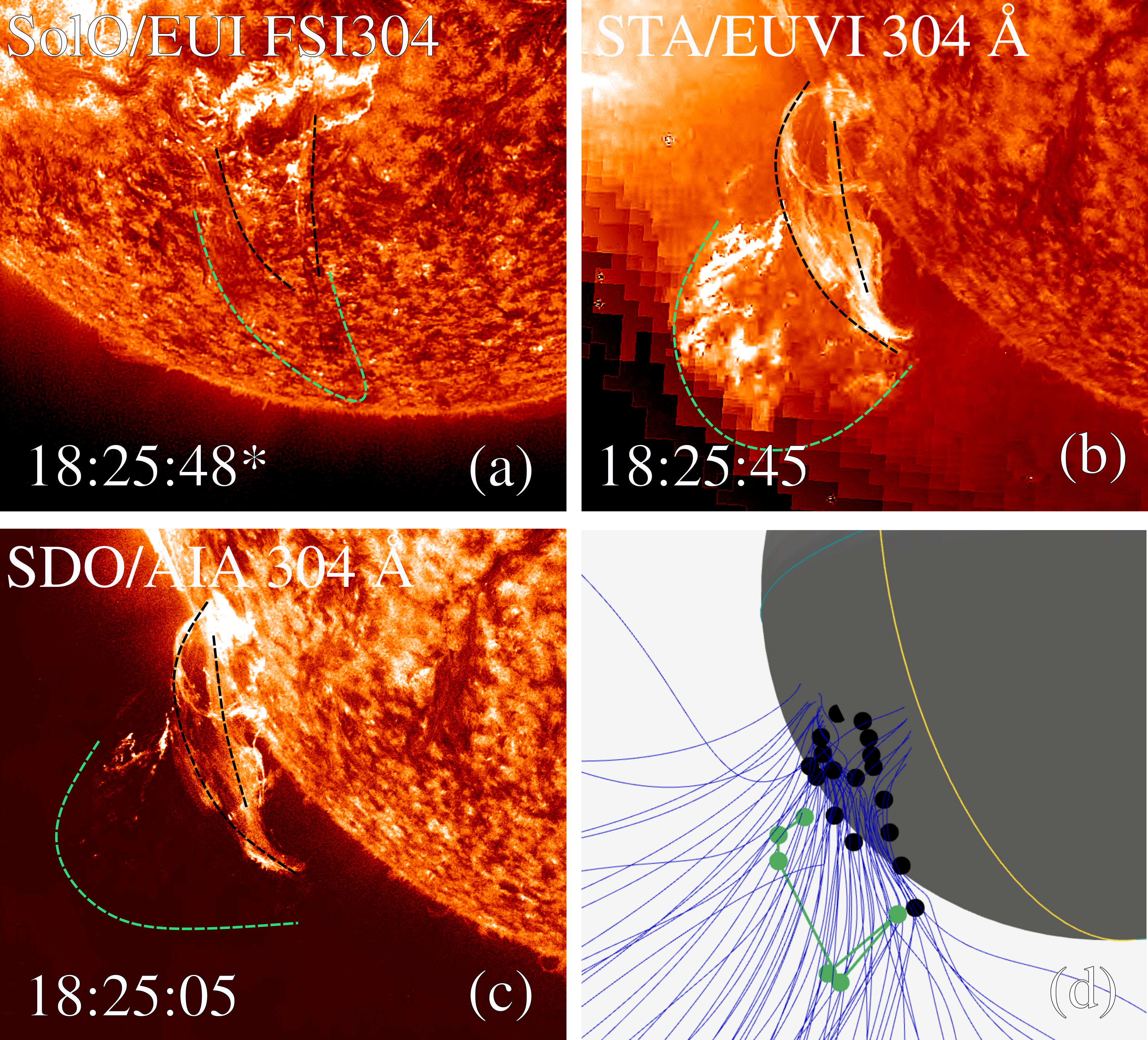} }
   
 \caption{Splitting filament on 2023 October 15 seen by SolO/EUI FSI304 (a), STA/EUVI 304 {\AA} (b), and SDO/AIA 304 {\AA} at 18:25 UT. SolO time corresponds to the time at Earth (\textit{date\_ear}) instead of the actual observation time, since the spacecraft received the image ahead of the other spacecraft. Schematically the reconstructed areas are indicated by lines projected in the different images. The green and black dashed lines indicate the eruptive and confined segments, respectively. In panel (d) the 3D position of the triangulated pixels are shown with the open field lines (blue) in the area of the eruption, and the colors of the segments as indicated above. An interactive version of this figure is available online, which allows different rotations and projections of the triangulated dots. The gray sphere represents the solar surface at $1\,R_\sun$, and to provide orientation the equator (teal), SolO meridian (yellow), STA meridian (pink), and SDO meridian (turquoise) are displayed.  } 
\label{fig:3d_c}
\end{figure}

Figure~\ref{fig:3d_b} shows an example of the reconstructed features by TP at 14:35 UT and 14:43 UT. At those times we identify a lower arc which corresponds to the confined segment (grey and black, respectively, line of each panel), and an upper arc corresponding to the eruptive segment (green lines). Panels (a) and (b) show the eruptive and confined segments at the same time from the SolO/EUI and SDO/AIA perspectives. Panel (c) shows a 304 \AA\ image at a later time from the SolO/EUI perspective. Darker green and black lines indicate the evolved eruptive and confined segments, respectively. Figure~\ref{fig:3d}(d) shows the 3D positions of the triangulated pixels and the closed field lines surrounding the area from the PFSS reconstruction. We note that the filament was rising above one lobe of the PS structure and that the eruptive segment moved towards the null point of this magnetic structure. Later in its evolution, the CME is deflected to the south (as the non-radial GCS parameters indicate), most likely by the negative-polarity coronal hole that overlaid the PS \citep[e.g., ][]{Sahade2023}. Meanwhile, the confined segment (grey and black dots) remained under the PS closed flux. An interactive version of Figure~\ref{fig:3d}(d) is available online, in which different angles of the 3D reconstruction can be displayed.

\subsection{2023 October 15}
A filament located in the southeast quadrant of the SolO/EUI disk  partially erupted after 17:55 UT. The eruptive filament formed with other filaments in the region days prior to the eruption. Checking the region days after the event and considering another filament located along the same filament channel, we infer that the east (west) leg of the filament was rooted in negative (positive) polarity.

The filament eruption was accompanied by footpoint brightenings observed in SolO/EUI 304 and 171 {\AA} images. At 18:15 UT, after reaching only $1.22\,R_\sun$, the eruption was partially halted and the splitting took place. The confined segment gradually dispersed, matching the position of different PFSS field lines, and remained visible until it drained back to the chromosphere around 20:00 UT. In the EUV images very little prominence material seems to be ejected, yet the eruptive segment produced a CME observed by both LASCO C2 and STA COR2. Figure~\ref{fig:2023-10-15} shows the evolution of this partial eruption from the SDO/AIA point of view, with a movie including all the filters and points of view available online.
In the top left panel the eruptive filament in the pre-splitting stage (white arrow) as observed in the 304 and 171 {\AA} filters; in the top right panel the erupting (green arrow) and confined (black arrow) segments are visible with prominence material connecting both segments. Then, in the bottom panel of Fig.~\ref{fig:2023-10-15}, the eruptive segment (green arrow) is observed simultaneously by SDO/AIA 304+171 and LASCO C2. The running-difference coronagraph image shows the CME, its shock, and the dark cavity. The CME front exhibits a speed of $\sim 500\,$km/s, obtained from the GCS reconstruction.
The PFSS reconstruction indicates that the filament was embedded in a PS with negative open field lines around it. The segmentation took place near the initial position of the PS null. The confined segment remained under the closed region while the eruptive one escaped, also producing a CME that deflected in the coronal hole orientation. Figure~\ref{fig:3d_c} shows the reconstructed features for the time 18:25 UT. At that time we identify a lower arc corresponding to the confined segment (black line of each panel), and an upper arc corresponding to the eruptive segment (green lines). Figure~\ref{fig:3d}(d) shows the 3D positions of the triangulated pixels and the open field lines surrounding the source region. We note that the confined segment, after the splitting, relocated  to the south, within the closed field lines of the PS.  An interactive version of panel (d) is available online, in which different views of the 3D reconstruction can be displayed.

\subsection{Kinematics}
\begin{figure*}    %%%%%%%%%%%%%%%%%% FIGURE 1 
   \centerline{\includegraphics[width=0.95\textwidth]{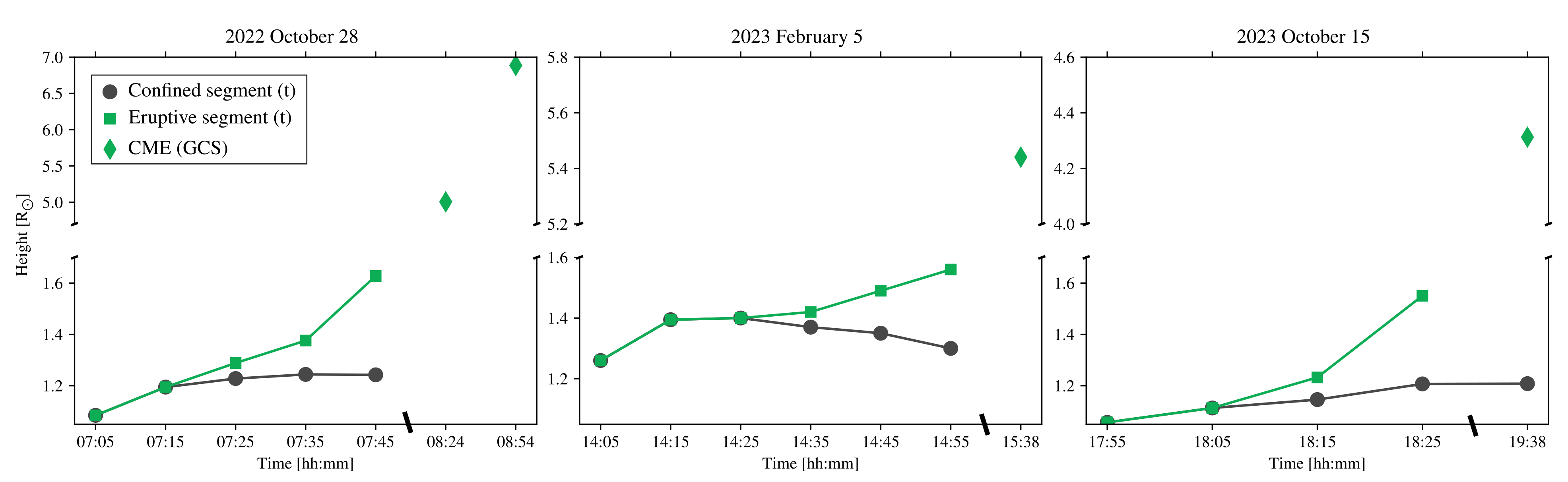}
             }
 \caption{Height vs. time for the three events. Black and green dots indicate the position of the confined and eruptive segments of the filament, respectively, as obtained with the \texttt{scc\_measure3} tool. Superimposed dots representing the confined and eruptive segments refer to their pre-splitting time. The green diamond markers indicate the height of the center of the cross section of the CME modeled with GCS. } 
\label{fig:kinem}
\end{figure*}
We present here three well-observed partial filament eruptions in which the filaments undergo splitting after they start rising. By reconstructing the real 3D position of different features of the prominence material, we assure the connection (and later disconnection) between segments of the filaments. The ejected portions of these events were associated with CMEs that we fit with the GCS model. Figure~\ref{fig:kinem} shows the evolution of height over time for the confined (black dots with solid line) and eruptive (green dots with solid line) segments. The dots correspond to the higher feature triangulated with \texttt{scc\_measure3} at each time, and the diamonds show the central height of the cross section of the GCS reconstruction of the CME. The three CMEs exhibit deflections aligned with the coronal-hole flux orientation, according to the non-radial GCS reconstruction and open field lines' orientation. 

All of the studied events originated in PSs. From the PFSS reconstructions we estimated the PS null heights. The first (2022-Oct-28) and third (2023-Oct-15) events were embedded in small PSs with null corridors located about $1.1\,R_\sun$.  The 2023 February PS had a higher null point at $\sim 1.5\,R_\sun$, connected to a helmet streamer towards the south. These estimates can be affected by the method of reconstruction and by the out-of-date magnetograms. As mentioned before, the filaments apparently split when their magnetic fields interacted with the external field at null points, as shown by the 3D position of the confined segments and the separation height with respect to the magnetic field reconstruction. After the splitting, the ejected segment produced CMEs in the range of $[500, 900]\,$km/s. The confined segments stopped rising or even fell until they were no longer visible.

\section{Discussion}
\label{sec:discussion}

To our knowledge, the observations presented above are the first showing partial eruptions that suffer segmentation after the onset of the eruption and, yet, also produce moderately fast and structured CMEs indicating the presence of an escaping MFR. These events are in contrast to the many previously reported partial eruptions leading to jets \citep[e.g.,][]{Mason2021,Zhang2022,Sun2023}. The events reported here represent a new class of partial filament eruptions. Unveiling the magnetic nature of this kind of partial eruptions, however, is not straightforward.

An alternative explanation of the observations presented here would be two filament eruptions that appear to be a single eruption at first but then separate; one escapes to form a CME, while the other fails and eventually drains back to the surface. We considered this alternative carefully, and found the following reasons for concluding it is not viable. Basically, the spatial, viewpoint, and temporal requirements for this scenario are so constrained that their occurrence is highly unlikely. We tracked the real 3D positions of the filament segments with the new routine  \texttt{scc\_measure3}, which provides more confidence in the reconstruction and ensures spatial connectivity between the segments.
For the 2023 October 15 event, which occurred on the disk as viewed by SolO, there is no evidence for a second filament either before or during the eruption, or for more than 2 flare ribbons (see animation accompanying Fig. 5). For the 2 limb events it is impossible to determine whether two filaments were present in the source region prior to eruption. Two eruptions from different segments of the same PIL would not follow the same path from initiation to the observed coronal heights, nor would they erupt at the exact same time. Sympathetic eruptions occur sequentially, as the overlying flux that is removed to allow one side to erupt must then be removed to enable the other side to erupt \citep[e.g., ][]{Zhou2021}. For a nearby filament on a different PIL to erupt without a visible gap between it and another filament, the conditions are even more difficult to meet: the relative timing must be precisely tuned, the two filaments must jointly appear to comprise a single feature, and their speeds must be very well matched. Moreover these conditions must be met from all 3 viewpoints. Therefore, we contend that the observed events are best explained by single flux ropes or sheared arcades that split in the corona.

We reconstructed the background magnetic field surrounding the eruption source region. Given the locations of the events, the photospheric magnetic field used for the PFSS model necessarily came from synoptic magnetograms a few days earlier or later than the eruption. This introduces some uncertainties, but the general topology of the region was stable during the Carrington rotation.
The PS topology common to all three events, along with reconnection signatures such as multi-wavelength brightenings and ribbons, suggests magnetic breakout  \citep{Antiochos1999} playing a key role in the eruptions and their partitioning. However, the fact that the upper part of the magnetic system was able to escape into the heliosphere, forming a CME, while the lower segment remained confined in a new magnetic equilibrium poses a challenge to our understanding.
\citet{Kumar2021} observed that a broad continuum of ejecta, from narrow jets to broad CMEs, can originate from filament channels in pseudostreamers.  In agreement with the observations, simulations of PS eruptions demonstrate that such events can produce true CMEs with closed flux extending to at least the inner heliosphere \citep{Wyper2024}. Figure 5 from the \citet{Wyper2024} shows a CME flux rope extending out to almost 10 solar radii before it disconnects from the Sun by interchange reconnection at the PS null point low in the corona. The eruptive part of our events might exhibit similar behavior, but the initial evolution of the three cases is difficult to understand from the theoretical viewpoint.

We consider the four standard models for eruption onset: kink instability, torus/loss-of-equilibrium, tether-cutting, and breakout \citep[e.g.,][]{Forbes2006,Chen2011}. In the kink instability model, the filament channel contains twisted flux that undergoes helical and rotational deformation leading to eruption onset.  In our observations, however, the rising filament does not twist or writhe; rather, it appears to rise straight up, which implies that the kink instability is not playing a role.  With regard to the torus instability/loss of equilibrium, first note that, as mentioned before, there is no evidence in these events for a double filament in which one flux tube lies above the other, as has been invoked for similar events \citep{Liu2012}.  The torus instability might account for failed eruptions if the initial filament-carrying flux rope were lying in a region with a torus-unstable decay index, but then as the rope rises it reaches a region with torus-stable decay index \citep[e.g.,][]{Luo2022}. In our cases, however, the region would have to be both stable and unstable, because part of the rope erupts while the other stays behind.  It seems unlikely that a torus instability/loss of equilibrium can account for the bifurcated events described above. 

Our observations are also difficult to reconcile with the reconnection-driven onset models. In the tether-cutting reconnection model \citep{Moore2001}, slow reconnection is assumed to begin inside the filament channel. This leads to the creation of a twisted flux rope that starts rising slowly, inducing more reconnection. This tether-cutting reconnection eventually transitions to a fast state leading to an explosive eruption \citep[e.g.,][]{Jiang2021}. In our observations the filaments initially rise, then stop, then the vertical splitting produces a stable filament and a CME. 
To explain our observations, tether-cutting would have to start underneath the rising filament and then stop, and a new separate round of flare reconnection would have to start at a different flux surface inside the rope. Although it cannot be ruled out, such a scenario seems contrived and certainly has never been demonstrated.

As mentioned above, the overlying PSs common to the three studied eruptions present a favorable scenario for the breakout mechanism. The observed initial rise and subsequent evolution are most easily understood with a sheared-arcade topology for the filament channel. In this case the initial rise would be due to pure breakout reconnection at the distorted PS null between the closed field overlying the sheared arcade and the external open flux, with no reconnection below the observed prominence arch, so that the filament-channel field retains a sheared arcade structure throughout the initial rise. In this way, the eruptive flux rope will form later, as the upward rise continues, once a current sheet forms inside the filament channel. Tether-cutting could start there, eventually transitioning to flare reconnection and a fast eruption as in the standard models \citep{Karpen2012}. The observed lack of twist in the non-erupting portions of these events provides additional support for the sheared-arcade hypothesis, because this portion of the internal PS flux would be below the flare reconnection site.  The question remains, however, as to why the sheared arcade halts its initial rise and whether the PS null plays a role in this hiatus. To our knowledge, no numerical simulation using any onset mechanism, including breakout,  has shown the abrupt stop of a prominence rise followed by an eruptive splitting, as observed.

%Having discarded other mechanisms, we describe in the next section what we contend to be the most reasonable explanation for these observations.

\section{Conclusions}
\label{sec:conclusions}

Our observations pose a major challenge for all the standard eruption models: the rising phase is followed by halting of the lower part of the eruption. The halting and splitting seem to occur around the initial height of the PS null, but it is important to note that the null undoubtedly distorts and moves in response to the rising flux rope and breakout reconnection.  As considered in the previous section, none of the eruptive mechanisms alone seems to explain the observations. Instead perhaps a combination of mechanisms is the key.
For the 2023 October 15 event in particular, the observation of remote and parallel ribbons (see animation accompanying Fig. 5 about 18:00 and 18:12 UT) indicate breakout and tether-cutting reconnection, respectively \citep{Joshi2017}. However, the viewpoints and cadence of SolO are not good enough to resolve the reconnection timing and then confirm which is the dominant mechanism at each step of the eruption.  Further observations are clearly needed in order to answer this outstanding question. It would be particularly illuminating to have multi-vantage-point observations of both the motions and the underlying flare ribbons, if any. If we could determine precisely the location and timing of any reconnection-related heating associated with this type of event, we would be much better positioned to understand their physical origin. In addition, updated observations of the photospheric magnetic field could help to understand the detailed coronal dynamics.  
 
The partial eruptions reported here represent a new element in the continuum of eruptive events, from confined to ``full'' fast CMEs.
Solar eruptions release solar material and magnetic fields into the interplanetary medium. They can produce strong disturbances and geomagnetic storms, so it is of utmost importance to improve our predictive capabilities for these events. A crucial step toward this goal is to understand what determines how much magnetic flux and plasma is expelled during an eruption, in other words, how much of the free energy stored in the pre-eruption corona is actually released. The partial MFR eruptions are worthwhile to study in this context.
To unveil their magnetic nature we need to analyze more examples, preferably with coincident photospheric magnetic-field information, and to simulate eruptions exploring more flexible configurations that could lead to this puzzling phenomenon.

\begin{acknowledgements}
    AS was supported by an appointment to the NASA Postdoctoral Program at the NASA Goddard Space Flight Center, administered by Oak Ridge Associated Universities under contract with NASA. SKA was supported by a NASA LWS grant and a NSF SHINE grant to the University of Michigan. JTK was supported by a NASA LWS grant and GSFC ISFM funding. 
    The authors acknowledge the use of Solo/EUI, SDO/AIA, SOHO/LASCO, and STEREO/EUVI, COR1 and COR2 data. Animations from the observations were produced with JHelioviewer \citep{JHelioviewer}.
    The authors also wish to acknowledge and thank Emily Mason and an anonymous reviewer for constructive feedback on the manuscript.
 \end{acknowledgements}

\bibliography{biblio}{}

\begin{thebibliography}{}
\expandafter\ifx\csname natexlab\endcsname\relax\def\natexlab#1{#1}\fi
\providecommand{\url}[1]{\href{#1}{#1}}
\providecommand{\dodoi}[1]{doi:~\href{http://doi.org/#1}{\nolinkurl{#1}}}
\providecommand{\doeprint}[1]{\href{http://ascl.net/#1}{\nolinkurl{http://ascl.net/#1}}}
\providecommand{\doarXiv}[1]{\href{https://arxiv.org/abs/#1}{\nolinkurl{https://arxiv.org/abs/#1}}}

\bibitem[{{Antiochos}(1998)}]{Antiochos1998}
{Antiochos}, S.~K. 1998, \apjl, 502, L181, \dodoi{10.1086/311507}

\bibitem[{{Antiochos} {et~al.}(1999){Antiochos}, {DeVore}, \& {Klimchuk}}]{Antiochos1999}
{Antiochos}, S.~K., {DeVore}, C.~R., \& {Klimchuk}, J.~A. 1999, \apj, 510, 485, \dodoi{10.1086/306563}

\bibitem[{{Bi} {et~al.}(2015){Bi}, {Jiang}, {Yang}, {Xiang}, {Cai}, \& {Liu}}]{Bi2015}
{Bi}, Y., {Jiang}, Y., {Yang}, J., {et~al.} 2015, \apj, 805, 48, \dodoi{10.1088/0004-637X/805/1/48}

\bibitem[{{Brueckner} {et~al.}(1995){Brueckner}, {Howard}, {Koomen}, {Korendyke}, {Michels}, {Moses}, {Socker}, {Dere}, {Lamy}, {Llebaria}, {Bout}, {Schwenn}, {Simnett}, {Bedford}, \& {Eyles}}]{LASCO_1995SoPh}
{Brueckner}, G.~E., {Howard}, R.~A., {Koomen}, M.~J., {et~al.} 1995, \solphys, 162, 357, \dodoi{10.1007/BF00733434}

\bibitem[{{Chen} {et~al.}(2021){Chen}, {Su}, {Liu}, {Kliem}, {Zhang}, {Ji}, \& {Liu}}]{Chen2021ApJ}
{Chen}, J., {Su}, Y., {Liu}, R., {et~al.} 2021, \apj, 923, 142, \dodoi{10.3847/1538-4357/ac2ba1}

\bibitem[{{Chen}(2011)}]{Chen2011}
{Chen}, P.~F. 2011, Living Reviews in Solar Physics, 8, 1, \dodoi{10.12942/lrsp-2011-1}

\bibitem[{{Cheng} {et~al.}(2014){Cheng}, {Ding}, {Zhang}, {Sun}, {Guo}, {Wang}, {Kliem}, \& {Deng}}]{Cheng2014ApJ}
{Cheng}, X., {Ding}, M.~D., {Zhang}, J., {et~al.} 2014, \apj, 789, 93, \dodoi{10.1088/0004-637X/789/2/93}

\bibitem[{{Cheng} {et~al.}(2018){Cheng}, {Kliem}, \& {Ding}}]{Cheng2018}
{Cheng}, X., {Kliem}, B., \& {Ding}, M.~D. 2018, \apj, 856, 48, \dodoi{10.3847/1538-4357/aab08d}

\bibitem[{{Dumbovi{\'c}} {et~al.}(2024){Dumbovi{\'c}}, {Karbonini}, {{\v{C}}alogovi{\'c}}, {Matkovi{\'c}}, {Martini{\'c}}, {Remeshan}, {Braj{\v{s}}a}, \& {Vr{\v{s}}nak}}]{Mateja2024}
{Dumbovi{\'c}}, M., {Karbonini}, L., {{\v{C}}alogovi{\'c}}, J., {et~al.} 2024, \solphys, 299, 66, \dodoi{10.1007/s11207-024-02304-z}

\bibitem[{{Filippov}(2019)}]{Filippov2019}
{Filippov}, B.~P. 2019, Physics Uspekhi, 62, 847, \dodoi{10.3367/UFNe.2018.10.038467}

\bibitem[{{Forbes} {et~al.}(2006){Forbes}, {Linker}, {Chen}, {Cid}, {K{\'o}ta}, {Lee}, {Mann}, {Miki{\'c}}, {Potgieter}, {Schmidt}, {Siscoe}, {Vainio}, {Antiochos}, \& {Riley}}]{Forbes2006}
{Forbes}, T.~G., {Linker}, J.~A., {Chen}, J., {et~al.} 2006, \ssr, 123, 251, \dodoi{10.1007/s11214-006-9019-8}

\bibitem[{{Gilbert} {et~al.}(2000){Gilbert}, {Holzer}, {Burkepile}, \& {Hundhausen}}]{Gilbert2000}
{Gilbert}, H.~R., {Holzer}, T.~E., {Burkepile}, J.~T., \& {Hundhausen}, A.~J. 2000, \apj, 537, 503, \dodoi{10.1086/309030}

\bibitem[{{Green} {et~al.}(2018){Green}, {T{\"o}r{\"o}k}, {Vr{\v{s}}nak}, {Manchester}, \& {Veronig}}]{Green2018}
{Green}, L.~M., {T{\"o}r{\"o}k}, T., {Vr{\v{s}}nak}, B., {Manchester}, W., \& {Veronig}, A. 2018, \ssr, 214, 46, \dodoi{10.1007/s11214-017-0462-5}

\bibitem[{{Hou} {et~al.}(2023){Hou}, {Li}, {Li}, {Su}, {Qiu}, {Yang}, {Yang}, {Li}, {Guo}, {Hou}, {Song}, {Bai}, {Zhou}, {Ding}, {Gan}, \& {Deng}}]{Hou2023}
{Hou}, Y., {Li}, C., {Li}, T., {et~al.} 2023, \apj, 959, 69, \dodoi{10.3847/1538-4357/ad08bd}

\bibitem[{{Howard} {et~al.}(2008){Howard}, {Moses}, {Vourlidas}, {Newmark}, {Socker}, {Plunkett}, {Korendyke}, {Cook}, {Hurley}, {Davila}, \& {et al.}}]{SECCHI_2008}
{Howard}, R.~A., {Moses}, J.~D., {Vourlidas}, A., {et~al.} 2008, \ssr, 136, 67, \dodoi{10.1007/s11214-008-9341-4}

\bibitem[{{Inhester}(2006)}]{Inhester2006}
{Inhester}, B. 2006, arXiv Astrophysics e-prints

\bibitem[{{Jiang} {et~al.}(2018){Jiang}, {Feng}, \& {Hu}}]{Jiang2018}
{Jiang}, C., {Feng}, X., \& {Hu}, Q. 2018, \apj, 866, 96, \dodoi{10.3847/1538-4357/aadd08}

\bibitem[{{Jiang} {et~al.}(2021){Jiang}, {Feng}, {Liu}, {Yan}, {Hu}, {Moore}, {Duan}, {Cui}, {Zuo}, {Wang}, \& {Wei}}]{Jiang2021}
{Jiang}, C., {Feng}, X., {Liu}, R., {et~al.} 2021, Nature Astronomy, 5, 1126, \dodoi{10.1038/s41550-021-01414-z}

\bibitem[{{Joshi} {et~al.}(2017){Joshi}, {Sterling}, {Moore}, {Magara}, \& {Moon}}]{Joshi2017}
{Joshi}, N.~C., {Sterling}, A.~C., {Moore}, R.~L., {Magara}, T., \& {Moon}, Y.-J. 2017, \apj, 845, 26, \dodoi{10.3847/1538-4357/aa7c1b}

\bibitem[{{Kaiser} {et~al.}(2008){Kaiser}, {Kucera}, {Davila}, {St.~Cyr}, {Guhathakurta}, \& {Christian}}]{STEREO_2008}
{Kaiser}, M.~L., {Kucera}, T.~A., {Davila}, J.~M., {et~al.} 2008, Space Sci. Rev., 136, 5, \dodoi{10.1007/s11214-007-9277-0}

\bibitem[{{Kang} {et~al.}(2023){Kang}, {Guo}, {Li}, {Wang}, \& {Lin}}]{Kang2023}
{Kang}, K., {Guo}, Y., {Li}, Y., {Wang}, J., \& {Lin}, J. 2023, Research in Astronomy and Astrophysics, 23, 095018, \dodoi{10.1088/1674-4527/ace519}

\bibitem[{{Karpen} {et~al.}(2012){Karpen}, {Antiochos}, \& {DeVore}}]{Karpen2012}
{Karpen}, J.~T., {Antiochos}, S.~K., \& {DeVore}, C.~R. 2012, \apj, 760, 81, \dodoi{10.1088/0004-637X/760/1/81}

\bibitem[{{Kliem} {et~al.}(2014){Kliem}, {T{\"o}r{\"o}k}, {Titov}, {Lionello}, {Linker}, {Liu}, {Liu}, \& {Wang}}]{Kliem2014}
{Kliem}, B., {T{\"o}r{\"o}k}, T., {Titov}, V.~S., {et~al.} 2014, \apj, 792, 107, \dodoi{10.1088/0004-637X/792/2/107}

\bibitem[{{Kumar} {et~al.}(2021){Kumar}, {Karpen}, {Antiochos}, {Wyper}, {DeVore}, \& {Lynch}}]{Kumar2021}
{Kumar}, P., {Karpen}, J.~T., {Antiochos}, S.~K., {et~al.} 2021, \apj, 907, 41, \dodoi{10.3847/1538-4357/abca8b}

\bibitem[{{Lemen} {et~al.}(2012){Lemen}, {Title}, {Akin}, {Boerner}, {Chou}, {Drake}, {Duncan}, {Edwards}, {Friedlaender}, {Heyman}, \& {et al.}}]{AIA_2012SoPh}
{Lemen}, J.~R., {Title}, A.~M., {Akin}, D.~J., {et~al.} 2012, \solphys, 275, 17, \dodoi{10.1007/s11207-011-9776-8}

\bibitem[{{Li} {et~al.}(2022){Li}, {Peter}, {Chitta}, {Song}, {Xu}, \& {Xiang}}]{Li2022}
{Li}, L., {Peter}, H., {Chitta}, L.~P., {et~al.} 2022, \apj, 935, 85, \dodoi{10.3847/1538-4357/ac7ffa}

\bibitem[{{Liu} {et~al.}(2007){Liu}, {Alexander}, \& {Gilbert}}]{Liu2007}
{Liu}, R., {Alexander}, D., \& {Gilbert}, H.~R. 2007, \apj, 661, 1260, \dodoi{10.1086/513269}

\bibitem[{{Liu} {et~al.}(2012){Liu}, {Kliem}, {T{\"o}r{\"o}k}, {Liu}, {Titov}, {Lionello}, {Linker}, \& {Wang}}]{Liu2012}
{Liu}, R., {Kliem}, B., {T{\"o}r{\"o}k}, T., {et~al.} 2012, \apj, 756, 59, \dodoi{10.1088/0004-637X/756/1/59}

\bibitem[{{Luo} \& {Liu}(2022)}]{Luo2022}
{Luo}, R., \& {Liu}, R. 2022, \apj, 929, 2, \dodoi{10.3847/1538-4357/ac5b06}

\bibitem[{{Martin}(1998)}]{Martin1998}
{Martin}, S.~F. 1998, \solphys, 182, 107, \dodoi{10.1023/A:1005026814076}

\bibitem[{{Mason} {et~al.}(2021){Mason}, {Antiochos}, \& {Vourlidas}}]{Mason2021}
{Mason}, E.~I., {Antiochos}, S.~K., \& {Vourlidas}, A. 2021, \apjl, 914, L8, \dodoi{10.3847/2041-8213/ac0259}

\bibitem[{{McCauley} {et~al.}(2015){McCauley}, {Su}, {Schanche}, {Evans}, {Su}, {McKillop}, \& {Reeves}}]{McCauley2015}
{McCauley}, P.~I., {Su}, Y.~N., {Schanche}, N., {et~al.} 2015, \solphys, 290, 1703, \dodoi{10.1007/s11207-015-0699-7}

\bibitem[{{Moore} {et~al.}(2001){Moore}, {Sterling}, {Hudson}, \& {Lemen}}]{Moore2001}
{Moore}, R.~L., {Sterling}, A.~C., {Hudson}, H.~S., \& {Lemen}, J.~R. 2001, \apj, 552, 833, \dodoi{10.1086/320559}

\bibitem[{{M{\"u}ller} {et~al.}(2017){M{\"u}ller}, {Nicula}, {Felix}, {Verstringe}, {Bourgoignie}, {Csillaghy}, {Berghmans}, {Jiggens}, {Garc{\'\i}a-Ortiz}, {Ireland}, {Zahniy}, \& {Fleck}}]{JHelioviewer}
{M{\"u}ller}, D., {Nicula}, B., {Felix}, S., {et~al.} 2017, \aap, 606, A10, \dodoi{10.1051/0004-6361/201730893}

\bibitem[{{M{\"u}ller} {et~al.}(2020){M{\"u}ller}, {St. Cyr}, {Zouganelis}, {Gilbert}, {Marsden}, {Nieves-Chinchilla}, {Antonucci}, {Auch{\`e}re}, {Berghmans}, {Horbury}, {Howard}, {Krucker}, {Maksimovic}, {Owen}, {Rochus}, {Rodriguez-Pacheco}, {Romoli}, {Solanki}, {Bruno}, {Carlsson}, {Fludra}, {Harra}, {Hassler}, {Livi}, {Louarn}, {Peter}, {Sch{\"u}hle}, {Teriaca}, {del Toro Iniesta}, {Wimmer-Schweingruber}, {Marsch}, {Velli}, {De Groof}, {Walsh}, \& {Williams}}]{SOLO_2020A&A}
{M{\"u}ller}, D., {St. Cyr}, O.~C., {Zouganelis}, I., {et~al.} 2020, \aap, 642, A1, \dodoi{10.1051/0004-6361/202038467}

\bibitem[{{Patsourakos} {et~al.}(2020){Patsourakos}, {Vourlidas}, {T{\"o}r{\"o}k}, {Kliem}, {Antiochos}, {Archontis}, {Aulanier}, {Cheng}, {Chintzoglou}, {Georgoulis}, {Green}, {Leake}, {Moore}, {Nindos}, {Syntelis}, {Yardley}, {Yurchyshyn}, \& {Zhang}}]{Patsourakos2020}
{Patsourakos}, S., {Vourlidas}, A., {T{\"o}r{\"o}k}, T., {et~al.} 2020, \ssr, 216, 131, \dodoi{10.1007/s11214-020-00757-9}

\bibitem[{{Pesnell} {et~al.}(2012){Pesnell}, {Thompson}, \& {Chamberlin}}]{SDO_2012SoPh}
{Pesnell}, W.~D., {Thompson}, B.~J., \& {Chamberlin}, P.~C. 2012, \solphys, 275, 3, \dodoi{10.1007/s11207-011-9841-3}

\bibitem[{{Raouafi} {et~al.}(2016){Raouafi}, {Patsourakos}, {Pariat}, {Young}, {Sterling}, {Savcheva}, {Shimojo}, {Moreno-Insertis}, {DeVore}, {Archontis}, {T{\"o}r{\"o}k}, {Mason}, {Curdt}, {Meyer}, {Dalmasse}, \& {Matsui}}]{Raouafi2016}
{Raouafi}, N.~E., {Patsourakos}, S., {Pariat}, E., {et~al.} 2016, \ssr, 201, 1, \dodoi{10.1007/s11214-016-0260-5}

\bibitem[{{Rochus} {et~al.}(2020){Rochus}, {Auch{\`e}re}, {Berghmans}, {Harra}, {Schmutz}, {Sch{\"u}hle}, {Addison}, {Appourchaux}, {Aznar Cuadrado}, {Baker}, {Barbay}, {Bates}, {BenMoussa}, {Bergmann}, {Beurthe}, {Borgo}, {Bonte}, {Bouzit}, {Bradley}, {B{\"u}chel}, {Buchlin}, {B{\"u}chner}, {Cab{\'e}}, {Cadiergues}, {Chaigneau}, {Chares}, {Choque Cortez}, {Coker}, {Condamin}, {Coumar}, {Curdt}, {Cutler}, {Davies}, {Davison}, {Defise}, {Del Zanna}, {Delmotte}, {Delouille}, {Dolla}, {Dumesnil}, {D{\"u}rig}, {Enge}, {Fran{\c{c}}ois}, {Fourmond}, {Gillis}, {Giordanengo}, {Gissot}, {Green}, {Guerreiro}, {Guilbaud}, {Gyo}, {Haberreiter}, {Hafiz}, {Hailey}, {Halain}, {Hansotte}, {Hecquet}, {Heerlein}, {Hellin}, {Hemsley}, {Hermans}, {Hervier}, {Hochedez}, {Houbrechts}, {Ihsan}, {Jacques}, {J{\'e}r{\^o}me}, {Jones}, {Kahle}, {Kennedy}, {Klaproth}, {Kolleck}, {Koller}, {Kotsialos}, {Kraaikamp}, {Langer}, {Lawrenson}, {Le Clech'}, {Lenaerts}, {Liebecq}, {Linder}, {Long}, {Mampaey}, {Markiewicz-Innes}, {Marquet},
  {Marsch}, {Matthews}, {Mazy}, {Mazzoli}, {Meining}, {Meltchakov}, {Mercier}, {Meyer}, {Monecke}, {Monfort}, {Morinaud}, {Moron}, {Mountney}, {M{\"u}ller}, {Nicula}, {Parenti}, {Peter}, {Pfiffner}, {Philippon}, {Phillips}, {Plesseria}, {Pylyser}, {Rabecki}, {Ravet-Krill}, {Rebellato}, {Renotte}, {Rodriguez}, {Roose}, {Rosin}, {Rossi}, {Roth}, {Rouesnel}, {Roulliay}, {Rousseau}, {Ruane}, {Scanlan}, {Schlatter}, {Seaton}, {Silliman}, {Smit}, {Smith}, {Solanki}, {Spescha}, {Spencer}, {Stegen}, {Stockman}, {Szwec}, {Tamiatto}, {Tandy}, {Teriaca}, {Theobald}, {Tychon}, {van Driel-Gesztelyi}, {Verbeeck}, {Vial}, {Werner}, {West}, {Westwood}, {Wiegelmann}, {Willis}, {Winter}, {Zerr}, {Zhang}, \& {Zhukov}}]{EUI_2020A&A}
{Rochus}, P., {Auch{\`e}re}, F., {Berghmans}, D., {et~al.} 2020, \aap, 642, A8, \dodoi{10.1051/0004-6361/201936663}

\bibitem[{Sahade(2024)}]{scc_measure3}
Sahade, A. 2024, scc\_measure 3,  Zenodo, \dodoi{10.5281/zenodo.13951841}

\bibitem[{{Sahade} {et~al.}(2023){Sahade}, {Vourlidas}, {Balmaceda}, \& {C{\'e}cere}}]{Sahade2023}
{Sahade}, A., {Vourlidas}, A., {Balmaceda}, L.~A., \& {C{\'e}cere}, M. 2023, \apj, 953, 150, \dodoi{10.3847/1538-4357/ace420}

\bibitem[{{Sahade} {et~al.}(2025){Sahade}, {Vourlidas}, \& {Mac Cormack}}]{Sahade2025}
{Sahade}, A., {Vourlidas}, A., \& {Mac Cormack}, C. 2025, \apj, 978, 41, \dodoi{10.3847/1538-4357/ad96ba}

\bibitem[{{Scherrer} {et~al.}(2012){Scherrer}, {Schou}, {Bush}, {Kosovichev}, {Bogart}, {Hoeksema}, {Liu}, {Duvall}, {Zhao}, {Title}, {Schrijver}, {Tarbell}, \& {Tomczyk}}]{HMI2012}
{Scherrer}, P.~H., {Schou}, J., {Bush}, R.~I., {et~al.} 2012, \solphys, 275, 207, \dodoi{10.1007/s11207-011-9834-2}

\bibitem[{{Schrijver} \& {De Rosa}(2003)}]{PFSS_2003SoPh}
{Schrijver}, C.~J., \& {De Rosa}, M.~L. 2003, \solphys, 212, 165, \dodoi{10.1023/A:1022908504100}

\bibitem[{{Stenborg} {et~al.}(2008){Stenborg}, {Vourlidas}, \& {Howard}}]{Stenborg2008}
{Stenborg}, G., {Vourlidas}, A., \& {Howard}, R.~A. 2008, \apj, 674, 1201, \dodoi{10.1086/525556}

\bibitem[{{Sun} {et~al.}(2023){Sun}, {Li}, {Tian}, {Hou}, {Hou}, {Chen}, {Bai}, \& {Deng}}]{Sun2023}
{Sun}, Z., {Li}, T., {Tian}, H., {et~al.} 2023, \apj, 953, 148, \dodoi{10.3847/1538-4357/ace5b1}

\bibitem[{{Tripathi} {et~al.}(2009){Tripathi}, {Gibson}, {Qiu}, {Fletcher}, {Liu}, {Gilbert}, \& {Mason}}]{Tripathi2009}
{Tripathi}, D., {Gibson}, S.~E., {Qiu}, J., {et~al.} 2009, \aap, 498, 295, \dodoi{10.1051/0004-6361/200809801}

\bibitem[{{van Driel-Gesztelyi} \& {Green}(2015)}]{vanDriel2015}
{van Driel-Gesztelyi}, L., \& {Green}, L.~M. 2015, Living Reviews in Solar Physics, 12, 1, \dodoi{10.1007/lrsp-2015-1}

\bibitem[{{Wang} {et~al.}(2015){Wang}, {Liu}, {Dai}, {Yang}, {Huang}, \& {Hu}}]{Wang2015}
{Wang}, R., {Liu}, Y.~D., {Dai}, X., {et~al.} 2015, \apj, 814, 80, \dodoi{10.1088/0004-637X/814/1/80}

\bibitem[{{Wyper} {et~al.}(2017){Wyper}, {Antiochos}, \& {DeVore}}]{Wyper2017}
{Wyper}, P.~F., {Antiochos}, S.~K., \& {DeVore}, C.~R. 2017, \nat, 544, 452, \dodoi{10.1038/nature22050}

\bibitem[{{Wyper} {et~al.}(2021){Wyper}, {Antiochos}, {DeVore}, {Lynch}, {Karpen}, \& {Kumar}}]{Wyper2021}
{Wyper}, P.~F., {Antiochos}, S.~K., {DeVore}, C.~R., {et~al.} 2021, \apj, 909, 54, \dodoi{10.3847/1538-4357/abd9ca}

\bibitem[{{Wyper} {et~al.}(2024){Wyper}, {Lynch}, {DeVore}, {Kumar}, {Antiochos}, \& {Daldorff}}]{Wyper2024}
{Wyper}, P.~F., {Lynch}, B.~J., {DeVore}, C.~R., {et~al.} 2024, \apj, 975, 168, \dodoi{10.3847/1538-4357/ad7941}

\bibitem[{{Yang} {et~al.}(2018){Yang}, {Dai}, {Chen}, {Li}, \& {Jiang}}]{Yang2018}
{Yang}, J., {Dai}, J., {Chen}, H., {Li}, H., \& {Jiang}, Y. 2018, \apj, 862, 86, \dodoi{10.3847/1538-4357/aaccfd}

\bibitem[{{Zhang} {et~al.}(2001){Zhang}, {Dere}, {Howard}, {Kundu}, \& {White}}]{Zhang2001}
{Zhang}, J., {Dere}, K.~P., {Howard}, R.~A., {Kundu}, M.~R., \& {White}, S.~M. 2001, \apj, 559, 452, \dodoi{10.1086/322405}

\bibitem[{{Zhang} {et~al.}(2022){Zhang}, {Zhang}, {Dai}, {Li}, \& {Ji}}]{Zhang2022}
{Zhang}, Y., {Zhang}, Q., {Dai}, J., {Li}, D., \& {Ji}, H. 2022, \solphys, 297, 138, \dodoi{10.1007/s11207-022-02072-8}

\bibitem[{Zhou {et~al.}(2021)Zhou, Shen, Zhou, Tang, Duan, \& Tan}]{Zhou2021}
Zhou, C., Shen, Y., Zhou, X., {et~al.} 2021, \apj, 923, 45, \dodoi{10.3847/1538-4357/ac28a0}

\bibitem[{{Zuccarello} {et~al.}(2009){Zuccarello}, {Romano}, {Farnik}, {Karlicky}, {Contarino}, {Battiato}, {Guglielmino}, {Comparato}, \& {Ugarte-Urra}}]{Zuccarello2009}
{Zuccarello}, F., {Romano}, P., {Farnik}, F., {et~al.} 2009, \aap, 493, 629, \dodoi{10.1051/0004-6361:200809887}

\end{thebibliography}
\bibliographystyle{aasjournal}

\end{document}